# Tunable metal-insulator transitions in bilayer graphene by thermal annealing

Gopinadhan Kalon, Young Jun Shin, and Hyunsoo Yang [a)]

*Department of Electrical and Computer Engineering, NUSNNI-Nanocore, National University of Singapore, Singapore 117576*

Tunable and highly reproducible metal-insulator transitions have been observed in bilayer graphene upon thermal annealing at 400 K under high vacuum conditions. Before annealing, the sample is metallic in the whole temperature regime of study. Upon annealing, the conductivity changes from metallic to that of an insulator and the transition temperature is a function of annealing time. The pristine metallic state can be reinstated by exposing to air thereby inducing changes in the electronic properties by adsorbing water vapor, which makes graphene a technologically promising material for sensor applications.

[a)] e-mail address: eleyang@nus.edu.sg

Graphene consists of one or a few monolayers of carbon atoms arranged in a hexagonal periodic lattice in two dimensions with potential applications in nanoelectronics due to its remarkable electronic properties such as very high carrier mobility values of ~ 10,000 $cm^2/(V\cdot s)$ at room temperature, which is 10 times higher than in silicon.[1, 2] However, graphene has no energy band gap which prevents them to be used as useful transistors. Recent studies show that patterning graphene into nanoribbons opens up a band gap[3] but precise control of the properties is a challenge. Metal-insulator (M-I) transitions are another method to lead graphene to be useful in memory and sensor applications. M-I transitions have been theoretically predicted[4-8] and experimentally reported in graphene by a few groups under different conditions such as the adsorption of $NO_2$ gas molecules,[9] percolation driven M-I transition in graphene nanoribbons due to inhomogeneous electron-hole puddle formation,[10] and gate-induced insulating state in bilayer graphene[11]. However, thermal annealing as a way to induce M-I transitions in graphene has not been explored so far, which is one of the technologically feasible methods.

In this Letter, we report M-I transitions in graphene by thermal annealing at 400 K under high vacuum conditions (< $10^{-5}$ Torr). The pristine metallic state is completely reversible when graphene is exposed to atmosphere. The transport studies indicates that the M-I transition is due to a combination of various effects such as removal of water vapor, shift in the charge neutrality point (CNP) upon thermal annealing, and change in the charged impurity-puddle formation conditions.

The bilayer graphene was prepared by micromechanical exfoliation of Kish graphite followed by a transfer to a highly p-doped Si substrate, which was covered by a layer of 300 nm thick $SiO_2$. Mechanically cleaved graphene was identified by an optical microscope and quantified by Raman spectrophotometer. For further details on the sample preparation and identification, we refer to our earlier works.[12, 13] Electrodes were patterned by optical lithography and Cr (5 nm)/Au (100 nm) was deposited by a thermal evaporator. Standard lift-off procedures were followed after the deposition (see Fig. 1(a)). Figure 1(b) shows the Raman spectrum of pristine graphene. The spectrum shows prominently the G peak along with a 2D peak. A Lorentian fit of a 2D peak in the inset of Fig. 1(b) estimates the full width at half maximum of 2D as 52 $cm^{-1}$, indicating that the graphene is bilayer. The sample also has a very weak D peak suggesting a slight disorder in the sample.

Electrical transport measurements have been carried out in a Hall bar geometry as a function of temperature ($T$) in the range of 2-300 K. Electrical resistivity $\rho_{xx}(T)$ have been calculated from the measured four-probe resistance $R$, using the relation $\rho_{xx} = (W/L)R$, where $W$ is the width and $L$ is the length of the sample. In all the measurements, no gate voltage has been applied to the sample. Figure 1(c) shows the resistivity ($\rho_{xx}$) both as a function of $T$ and cooling rate for a bilayer graphene sample. The sample has been cooled with a definite cooling rate from 300 to 2 K under high vacuum conditions. After that the sample has been warmed up to 300 K at the same rate of cooling, which is defined as one complete cycle. To understand the effect of cooling rate on the transport properties, it has been varied from 3 to 15 K/min in each cooling cycle. Figure 1(c) shows that the sample is metallic and also there is a strong dependency of $\rho_{xx}$ on the cooling rate. The cooling rate dependency of $\rho_{xx}$ suggests that the sample has slight disorder which could lead to a weak localization (WL) of the charged particles. WL is a quantum mechanical manifestation of the constructive interference of the carrier wave functions at low temperature.[14] The magneto-transport measurements give a clear evidence of WL which we discuss later. Following this experiment, the sample has been thermally annealed by heating at 400 K as a function of annealing time. Figure 1(b) shows $\rho_{xx}$ as a function of $T$ at different annealing time. In all these measurements, cooling and warming rate has been fixed at 12 K/min. After thermally annealing at 400 K for 10 s, the resistivity shows an upturn at a temperature of ~200 K when the sample is cooled from 400 to 2 K. The increase in resistivity with decrease in temperature from 300 K indicates insulating nature of the sample. It is known that vacuum annealing on graphene helps to remove the adsorbed water vapor as well as shift CNP to lower gate voltages.[15] Note that our pristine graphene has a high CNP > 30 V.[12] After warming the sample to 400 K and annealing for 30 s, the M-I transition has been shifted up to ~300 K. On further annealing at 400 K for 60, 120, and 180 s, the M-I transition has been shifted to high temperatures causing a complete insulating behavior at room temperature. This indicates that by controlling the thermal annealing time, one can effectively induce M-I transitions at a variable temperature.

To understand the mechanism of M-I transitions in graphene sample, Hall measurements have been carried out before and after thermal annealing for 180 s. Carrier concentration ($n$) and the type of carrier have been calculated from the magnitude and sign of the Hall coefficient $R_H$ respectively, by using the relationship $R_H = V_H/(I\times H)$, and $n = 1/(R_H e)$, where $V_H$ is the Hall voltage, $I$ is the current through the sample which was 100 $\mu$A in our sample, $e$ is the electron charge, and $H$ is the applied magnetic field normal to the sample plane. The Hall mobility has been extracted using the relation $\mu=1/(en\rho_{xx})$. Hall measurements have identified that majority of the carriers are holes mainly due to the adsorbed water molecules.[16] Figure 2(a) shows the hole concentration and mobility of pristine graphene as a function of $T$. The carrier concentration starts to decrease as the sample is cooled from 300 K and almost independent at low temperatures. Mobility of the pristine graphene sample increases initially when cooled from 300 K and attains a maximum value of 1100 $cm^2/(V\cdot s)$ at a temperature of 150 K. The increase in mobility is due to lesser amount of scattering from charged impurities (dopants) when the temperature is lowered.[17] The acoustic phonons of graphene usually play a role in controlling the mobility as the sample is cooled from 300 K, which may lead to an increase in mobility due to reduced lattice vibrations. But at low temperatures (< 100 K) mobility shows a decrease, which is a result of enhanced scattering due to quantum interference effect (WL), the details of which we discuss later. This also implies that the role of phonons in the scattering is partially suppressed as otherwise, mobility should have increased or remained constant when the temperature is decreased.[18] The mobility and carrier concentration are inversely proportional to each other, implying charged impurity scattering[17] as the dominant transport mechanism. Figure 2(b) shows the carrier concentration and mobility as a function $T$ after annealing at 400 K under high vacuum conditions for 180 s. The carrier concentration has decreased from a value of ~ $9\times10^{12}$ $cm^{-2}$ for pristine graphene to ~ $5\times10^{12}$ $cm^{-2}$ after annealing, which suggests that annealing removed most of the adsorbed water molecules, which act as dopants. Also, the magnitude of $n$ after annealing suggests that the CNP has shifted more close to the zero bias, but the CNP is still away from zero gate voltage. Interestingly, the carrier concentration shows an initial increase and then decreases slowly as the temperature decreases from 300 K. After annealing, the mobility has reduced drastically from 950 to 440 $cm^2/(V\cdot s)$ at 300 K. The mobility initially shows a drastic reduction as the sample is cooled from 300 K but attains a temperature independent value of 300 $cm^2/(V\cdot s)$ with slight variations.

To understand the transport mechanism furthermore, magnetoresistance (MR) measurements have been carried out at different temperatures. Here MR is defined as $MR = (\rho(H) - \rho(0)) / \rho(0)$ where $\rho(H)$ is the resistivity at a magnetic field $H$. Figure 3(a) shows the MR data as a function of $H$ at 2 K. Also the angle $\theta$ has been varied from 0° ($H$ is perpendicular to the film plane) to 90° (*H* is parallel to the plane). The sample shows a negative MR at low fields with a strong angle dependency. The negative MR is due to the suppression of WL by the applied magnetic field and WL is of two dimensional in nature, which explains the strong angle dependency of MR. WL has been reported in graphene by a few groups[19-21] and its origin is mainly attributed to inter-valley scattering due to point defects with a size of the order of lattice spacing. Figure 3(b) shows the MR data as a function of *H* at 10 K. The negative MR at low magnetic fields is now less pronounced at 10 K since the phase decoherence rate is a function of temperature. To verify further the role of charged impurity scattering, MR measurements have been carried out after annealing for 180 s. Figure 3(c) shows the MR data as a function of *H* and angle $\theta$ at 2 K. The sample shows a very weak negative MR at low magnetic fields in comparison to the one from a sample before annealing, suggesting that the carriers are less coherently back scattered due to an increase in disorder caused by a change in the distribution of inhomogeneous carrier density profile. Figure 3(d) shows the MR as a function of *H* at 10 K after annealing. The sample does not show any negative MR around zero fields. In all the MR measurements, a positive MR has seen at higher fields, which is attributed to be the result of orbital effect and the asymmetry in the MR values for positive and negative fields is mostly due to geometry effect rather than any intrinsic effect.

A *D* peak in the Raman spectrum, cooling rate dependency of $\rho_{xx}(T)$, and negative MR at low magnetic fields imply that slight disorder/carrier inhomogeneity is present in the pristine metallic sample. Transport measurements indicate that after annealing both $n$ and $\mu$ has decreased due to increase in disorder/carrier inhomogeneity. A recent theoretical study by Hwang *et al.*[22] discussed that the bilayer graphene may have strong charge density inhomogeneity which results in potential energy fluctuation leading to a coexisting metallic and insulating transport behavior. In our sample, a charged impurity induced electron-hole puddle formation is highly possible where the amount of charged impurity is controlled via annealing as well as exposure to moisture. Hwang *et al.*[22] mentioned that bilayer graphene is most favorable to have inhomogeneous electron-hole puddles as compared to single and multilayer graphene since the potential fluctuation strength is much weaker in the later case. Therefore, controlling the amount of adsorbed molecules along with carrier inhomogeneity, which favors an insulator transition, in the bilayer graphene is one of the easiest ways of tuning the M-I transition. When the insulating sample is again exposed to atmosphere, the sample turns into metallic which means the process is reversible and can be used for humidity sensor applications.

In summary, a tunable and highly reproducible metal-insulator transition has been observed in bilayer graphene upon thermal annealing at 400 K under high vacuum conditions. The pristine graphene is metallic in the whole temperature regime of study. Upon annealing, the conductivity change from metallic to that of an insulator and the M-I transition temperature is a function of annealing time. The metallic state can be reinstated by exposing to air thereby inducing changes in the electronic properties by adsorption of water vapor, which makes graphene a technologically promising material for sensor applications. The negative MR at low magnetic fields as well as a *D* peak in the Raman spectra indicate that the sample has slight disorder which is responsible for the observed M-I transitions.

References:

[1] K. S. Novoselov, et al., Science **315**, 1379 (2007).
[2] K. S. Novoselov, A. K. Geim, S. V. Morozov, D. Jiang, M. I. Katsnelson, I. V. Grigorieva, S. V. Dubonos, and A. A. Firsov, Nature **438**, 197 (2005).
[3] M. Y. Han, B. Ozyilmaz, Y. B. Zhang, and P. Kim, Phys. Rev. Lett. **98**, 206805 (2007).
[4] L. Zhang, Y. Zhang, M. Khodas, T. Valla, and I. A. Zaliznyak, Phys. Rev. Lett. **105**, 046804 (2010).
[5] S. Das Sarma, M. P. Lilly, E. H. Hwang, L. N. Pfeiffer, K. W. West, and J. L. Reno, Phys. Rev. Lett. **94**, 136401 (2005).
[6] V. Juriccaronicacute, I. F. Herbut, and G. W. Semenoff, Phys. Rev. B **80**, 081405 (2009).
[7] A. Bostwick, J. L. McChesney, K. V. Emtsev, T. Seyller, K. Horn, S. D. Kevan, and E. Rotenberg, Phys. Rev. Lett. **103**, 056404 (2009).
[8] O. V. Kibis, Phys. Rev. B **81**, 165433 (2010).
[9] S. Y. Zhou, D. A. Siegel, A. V. Fedorov, and A. Lanzara, Phys. Rev. Lett. **101**, 086402 (2008).
[10] S. Adam, S. Cho, M. S. Fuhrer, and S. Das Sarma, Phys. Rev. Lett. **101**, 046404 (2008).
[11] J. B. Oostinga, H. B. Heersche, X. Liu, A. F. Morpurgo, and L. M. K. Vandersypen, Nat. Mater. **7**, 151 (2008).
[12] Y. J. Shin, G. Kalon, J. Son, J. H. Kwon, J. Niu, C. S. Bhatia, G. Liang, and H. Yang, Appl. Phys. Lett. **97**, 252102 (2010).
[13] Y. J. Shin, Y. Wang, H. Huang, G. Kalon, A. T. Wee, Z. Shen, C. S. Bhatia, and H. Yang, Langmuir **26**, 3798 (2010).
[14] A. F. Morpurgo and F. Guinea, Phys. Rev. Lett. **97**, 196804 (2006).
[15] J. Moser, A. Barreiro, and A. Bachtold, Appl. Phys. Lett. **91**, 163513 (2007).
[16] F. Schedin, A. K. Geim, S. V. Morozov, E. W. Hill, P. Blake, M. I. Katsnelson, and K. S. Novoselov, Nat. Mater. **6**, 652 (2007).
[17] J. H. Chen, C. Jang, S. Adam, M. S. Fuhrer, E. D. Williams, and M. Ishigami, Nat. Phys. **4**, 377 (2008).
[18] J.-H. Chen, C. Jang, S. Xiao, M. Ishigami, and M. S. Fuhrer, Nat. Nano. **3**, 206 (2008).
[19] F. V. Tikhonenko, D. W. Horsell, R. V. Gorbachev, and A. K. Savchenko, Phys. Rev. Lett. **100**, 056802 (2008).
[20] E. McCann, K. Kechedzhi, V. I. Fal'ko, H. Suzuura, T. Ando, and B. L. Altshuler, Phys. Rev. Lett. **97**, 146805 (2006).
[21] R. V. Gorbachev, F. V. Tikhonenko, A. S. Mayorov, D. W. Horsell, and A. K. Savchenko, Phys. Rev. Lett. **98**, 176805 (2007).
[22] E. H. Hwang and S. Das Sarma, Phys. Rev. B **82**, 081409 (2010).

**FIGURES**

FIG. 1. (a) Optical micrograph of the patterned graphene device along with electrodes. Scale bar is 20 $\mu$m. (b) Raman spectrum of bilayer graphene. The inset in (b) shows the 2D peak along with a theoretical fit. (c) Plot of resistivity ($\rho_{xx}$) vs. temperature ($T$) before annealing at different rates of cooling. (d) Plot of resistivity ($\rho_{\mathrm{xx}}$) vs. temperature ($T$) after annealing at 400 K as a function of annealing time. An arrow in (d) is a guide to the eye about the transition temperature.

FIG. 2. (a) Plot of carrier concentration ($n$) and mobility ($\mu$) of pristine graphene before annealing as a function of temperature. (b) Plot of $n$ and $\mu$ after annealing for 180 s at 400 K as a function of temperature.

FIG. 3. The MR as a function of magnetic field and field angles before annealing at 2 K (a) and 10 K (b). Plot of the MR as a function of magnetic fields and field angles after annealing at 400 K for 180 s at 2 K (c) and 10 K (d).

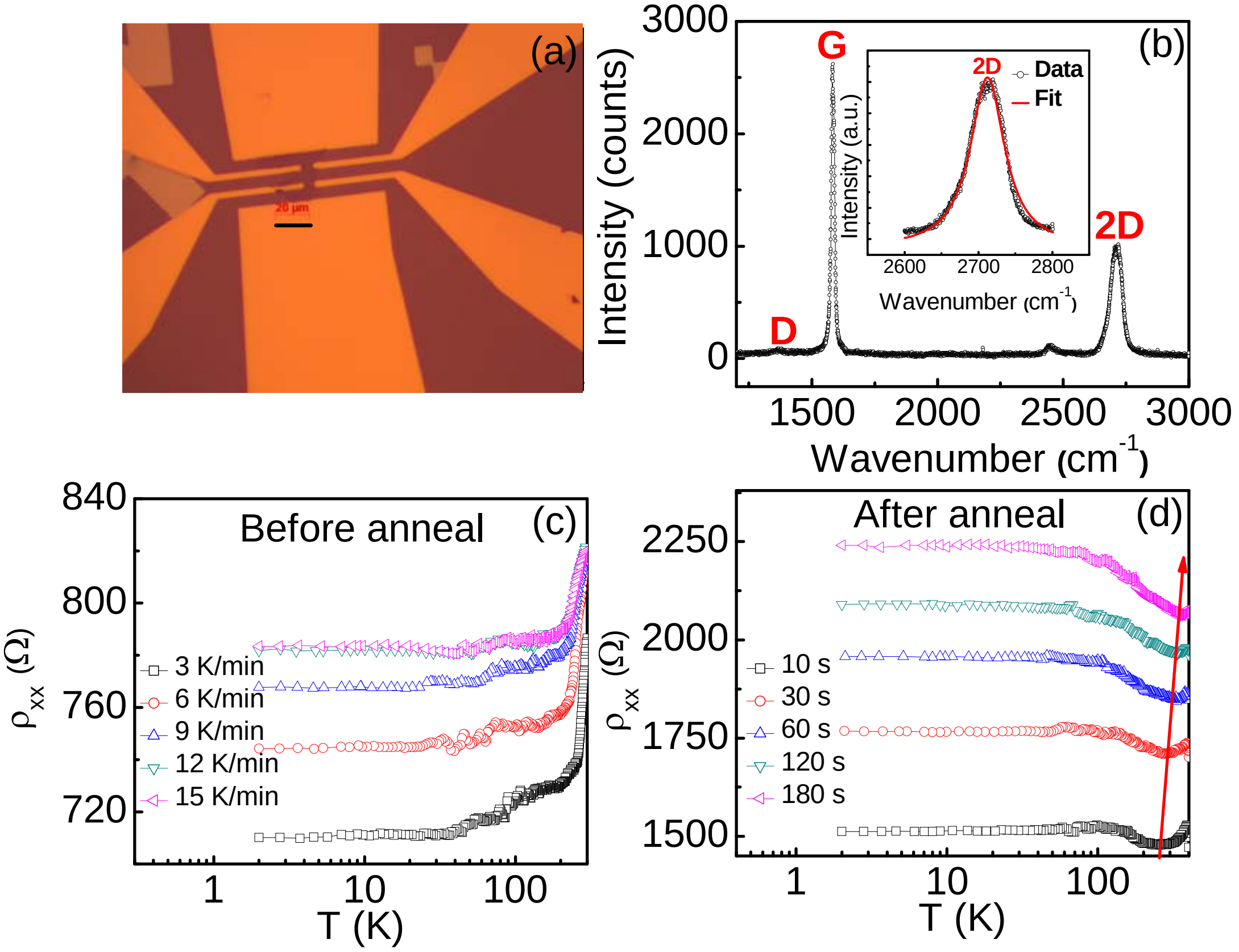

Figure 1

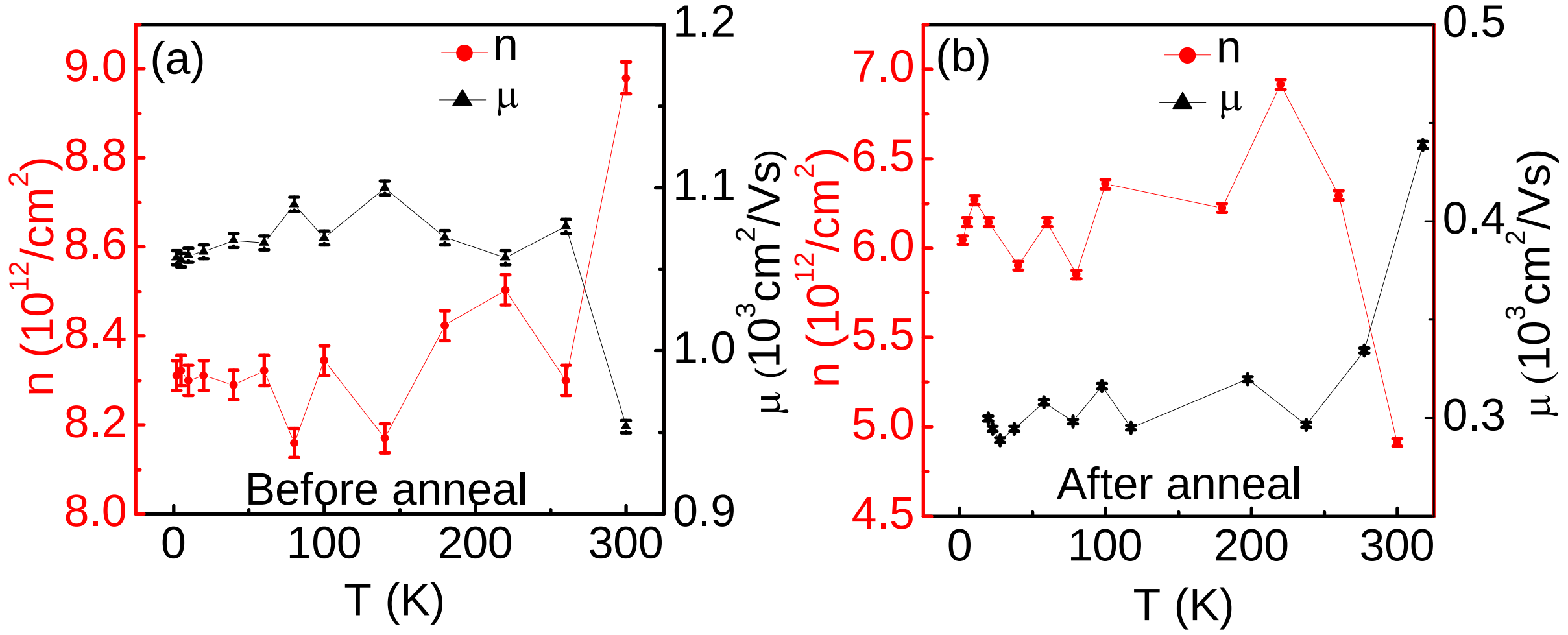

Figure 2

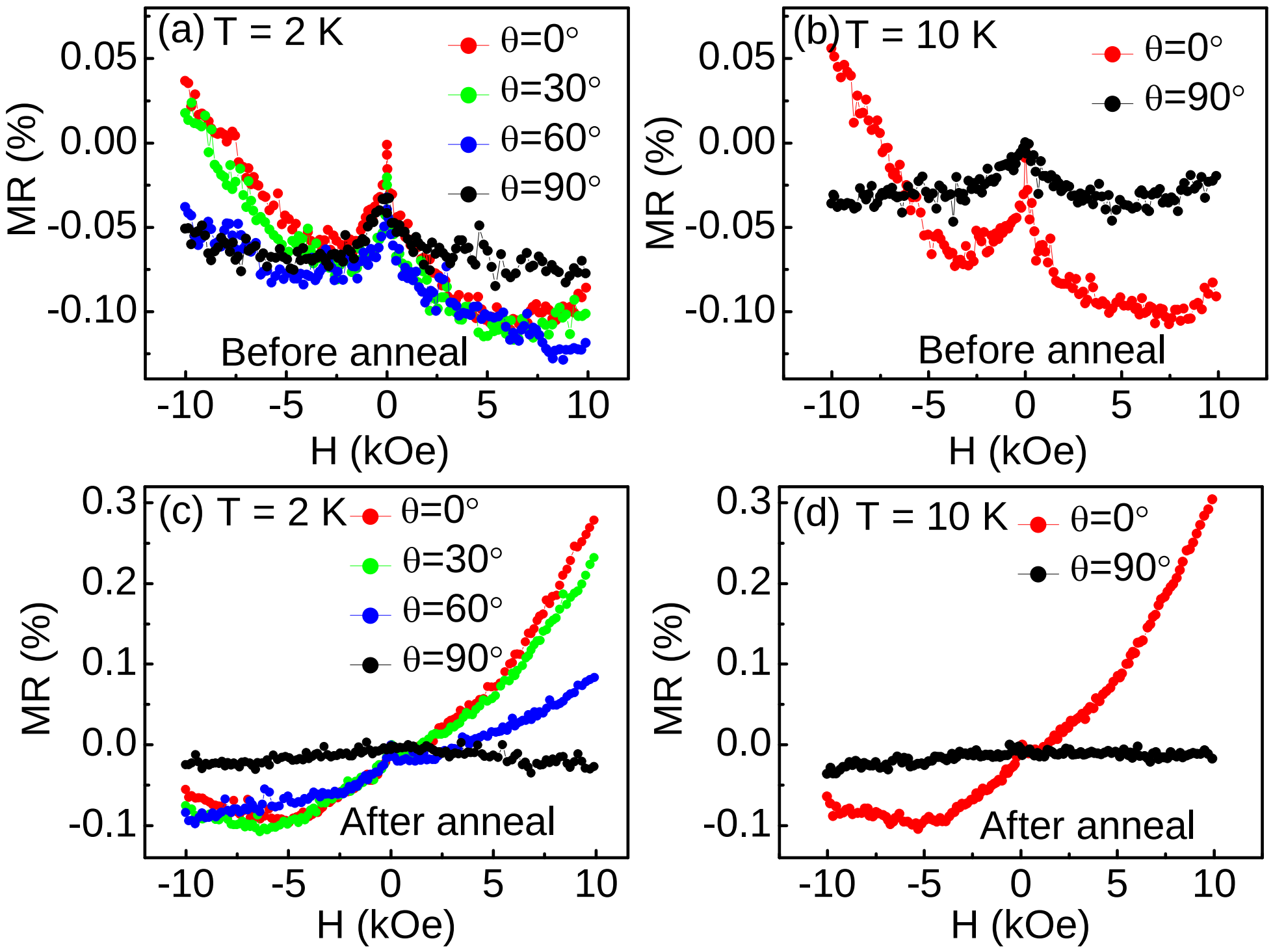

Figure 3